# A Rotating Collapsar and Possible Interpretation of the LSD Neutrino Signal from SN 1987A

V. S. Imshennik[1*] and O. G. Ryazhskaya[2]

[1]*Institute for Theoretical and Experimental Physics,*
*ul. Bol'shaya Cheremushkinskaya 25, Moscow, 117259 Russia*

[2]*Institute for Nuclear Research, Russian Academy of Sciences,*
*pr. Shestidesyatiletiya Oktyabrya 7a, Moscow, 117312 Russia*

**Abstract**—We consider an improved rotational mechanism of the explosion of a collapsing supernova. We show that this mechanism leads to two-stage collapse with a phase difference of ∼5 h. Based on this model, we attempt a new interpretation of the events in underground neutrino detectors on February 23, 1987, related to the supernova SN 1987A.

## INTRODUCTION

In effect, the idea of two-stage gravitational collapse has long been a subject of discussion, particularly in the case of the resumption (of the second stage) of the collapse in a neutron star with its transformation into a black hole (see, e.g., Imshennik and Nadyozhin 1988). However, if the rotation effects are taken into account, then such a two-stage collapse of an iron–oxygen–carbon (Fe–O–C) core acquires a different, more specific content that we have called the rotational mechanism of the explosion of a collapsing supernova (SN). The reason is that, in our opinion, the above rotation effects make it possible to solve the crucial problem of the transformation of the collapse into an explosion for high-mass and collapsing supernovae (all types of SN except the type-Ia thermonuclear SN). An extensive series of studies has been carried out since 1992 in connection with the famous SN 1987A (Imshennik and Nadyozhin 1992; Imshennik 1992; Aksenov and Imshennik 1994; Imshennik and Popov 1994; Aksenov *et al.* 1995; Imshennik 1995, 1996; Imshennik and Blinnikov 1996). Below, we consider the so-called improved rotational mechanism of explosion. The possibility of this mechanism was first mentioned by Imshennik and Popov (1994) in connection with the reception of two neutrino signals from SN 1987A (see Section 1) separated by a relatively long time interval of 4 h 44 min (≡4.7 h) (Dadykin *et al.* 1987; Aglietta *et al.* 1987) at $t_{UT}$ = 2 h 52 min (February 23, 1987) and $t_{UT}$ = 7 h 36 min (Hirata *et al.* 1987; Bionta *et al.* 1987). These observations of the neutrino signals were carefully analyzed by Dadykin *et al.* (1989). The above improvement of the explosion scenario for the rotating Fe–O–C core of a high-mass star stems not only from the necessity of explaining the observations of the neutrino signals from SN 1987A, but also from the intrinsic logic of the development of this scenario—it arose on the road to overcoming theoretical difficulties.

The neutrino spectra were obtained from theoretical estimates. These spectra are based on the hydrodynamic calculations of a quasi-one-dimensional model for the formation of a rotating collapsar (Imshennik and Nadyozhin 1977, 1992) and on the hypothesis of bulk neutrino radiation from a rotating collapsar (Imshennik and Nadyozhin 1972; Ivanova *et al.* 1969a) with the almost total dominance of electron neutrinos in the neutrino radiation (see Section 2). The derived spectra include the dimensionless chemical potential of electrons, $\varphi$, which is considered here as the only free parameter of these spectra.

In Section 3, we analyze the observations of the neutrino signal on the LSD detector at $t_{UT}$ = 2 h 52 min (February 23, 1987) by using the previously obtained neutrino spectra that additionally take into account the effects of self-absorption inside a rotating collapsar. We develop the hypothesis about the interaction of electron neutrinos with the nuclei of iron that is actually present in the LSD detector in large quantities. The products of these interaction reactions, mainly in the form of gamma-ray photons and electrons, are detected in a liquid scintillator with a photomultiplier. The detection efficiency is estimated by the Monte-Carlo method. We show that the observational data are consistent with the theory described in Sections 1 and 2 over a wide $\varphi$ range.

*E-mail: imshennik@itep.ru

## 1. THEORETICAL ANALYSIS OF THE TIME INTERVAL BETWEEN THE TWO STAGES OF COLLAPSE FOR SN 1987A

Thus, on the threshold of gravitational collapse, the Fe–O–C stellar core has a given (from calculations of the evolution of high-mass stars with a total mass on the main sequence $M_{\rm ms} \geq 10 M\odot$) mass $M_{\rm t}$ and a total angular momentum $J_0$, which are, obviously, conserved during the collapse of this core into a rotating collapsar. Aksenov *et al.* (1995) numerically constructed a large family of such two-dimensional, axisymmetric collapsars as a function of the parameters $M_{\rm t}$ and $J_0$. Dong and Shapiro (1995) proved the high probability of collapsars falling into the region of dynamical instability that is specified by the standard criterion $\beta = \mathcal{E}_{\rm rot}/|\mathcal{E}_{\rm grav}| \geq 0.27$ (Tassoul 1978). The quantities $\mathcal{E}_{\rm rot}$ and $|\mathcal{E}_{\rm grav}|$ denote the total rotational and total gravitational energies, respectively. Note that during collapse with the conservation of total angular momentum $J_0$ and local specific angular momentum, the energy $\mathcal{E}_{\rm rot}$ greatly increases compared to $|\mathcal{E}_{\rm grav}|$, which, of course, is also an increasing quantity. This instability grows with a characteristic hydrodynamic time and typically leads to the breakup of the collapsar into pieces, in the simplest case, into a binary of neutron stars (NS); almost all of the angular momentum can transform orbital angular momentum, $J_{\rm orb} \leq J_0$. However, $\Delta J = J_0 - J_{\rm orb}$ becomes the spin angular momentum of the NS themselves essentially in the more massive component of this binary. In other words, the NS binary is formed through the hydrodynamic fragmentation of a rotating collapsar. Imshennik (1992) showed that, under the additional assumption of a circular orbit of the binary and for given $M_{\rm t}$ and $J_{\rm orb}$, all of the orbital parameters, including the orbital radius $a_0$ and velocity $v_0$ (in terms of the reduced binary mass $M_1 M_2/M_{\rm t}$, according to the Kepler law) can be determined:

$$a_0 = \frac{J_{\rm orb}^2}{GM_{\rm t}^3} \cdot \frac{1}{\delta_0^2 (1-\delta_0)^2},$$

$$v_0 = \frac{GM_{\rm t}^2}{J_{\rm orb}} \cdot \delta_0 (1-\delta_0), \qquad \delta_0 = \frac{M_1}{M_{\rm t}}, \quad (1)$$

where the free parameter $0 \leq \delta_0 \leq 1/2$ appears. Below, $M_1$ is assumed to be the mass of the less massive NS. Remarkably, the evolution of the binary is determined by only one factor—gravitational radiation. The latter is unique from an astrophysical point of view, because it pertains to the evolution of such binaries, but in the presence of a low-mass NS whose mass is much lower than the mass of the more massive NS at the very outset; this mass is also small in absolute terms—compared to $M\odot$ (see below). In the point-mass approximation, the gravitational radiation (Peters and Mathews 1963; Landau and Lifshitz 1973) and the conservative mass transfer are described by a simple differential equation (Imshennik and Popov 2002):

$$\frac{da}{dt} = -\frac{64G^3 M_{\rm t}^3 \delta(1-\delta)}{5c^5 a^3} - 2a\frac{1-2\delta}{\delta(1-\delta)}\frac{d\delta}{dt}, \quad (2)$$

where, according to the common property of NS (degenerate stars), there is mass transfer with a decrease in the mass of the low-mass NS ($d\delta/dt < 0$). The derivative $d\delta/dt$ can be explicitly calculated in the Roche approximation (the Roche lobe and potential), as was shown in detail in the paper mentioned above, which continues the classic works in this field pioneered by Paczynski and Sienkiewicz (1972). Equation (2) then describes the evolution under the action of two factors; the second factor is mass transfer that causes the radius $a$ to increase, in contrast to the first factor that describes the approach of the components. However, the second factor comes into play only after the low-mass NS fills its Roche lobe. Before this time, the NS only approach each other at a constant parameter $\delta = \delta_0$ due to the gravitational radiation that carries away not only the energy of the NS binary, but also its orbital angular momentum. For $d\delta/dt = 0$, i.e., at $\delta = \delta_0$, Eq. (2) has a simple analytical solution from which $t_{\rm grav}$, the time of the closest approach of the components with a constant arbitrary parameter $\delta_0$, can be obtained formally up to the radius $a = 0$ (Imshennik and Popov 1994):

$$t_{\rm grav} = 2.94 \times 10^{-4} \frac{j_0^8}{m_{\rm t}^{15} \delta_0^9 (1-\delta_0)^9} \text{ s}, \quad (3)$$

where $j_0 = J_{\rm orb}/8.81 \times 10^{49}$ erg s and $m_{\rm t} = M_{\rm t}/2M\odot$; this choice of scales for the dimensionless quantities $j_0$ and $m_{\rm t}$ corresponds to the typical conditions of a Fe–O–C stellar core on the threshold of its collapse (as applied to SN 1987A). It should be noted that the arbitrariness in specifying the initial rotation in the stellar core is severely restricted by the hypothesis that this rotation is rigid; this is related to the action of convection inherent in the structure of high-mass stars, particularly at the final stages of their thermonuclear evolution. For this reason, the rotational energy in the initial conditions is actually negligible compared to the gravitational energy, i.e., $\beta \ll 1$, and the stellar structure is virtually spherically symmetric.

Table 1 presents the values of $\delta_0$ that follow from relation (3), as long as $m_{\rm t} = 0.9$ ($M_{\rm t} = 1.8 M\odot$) and, most importantly, as applied to SN 1987A, $t_{\rm grav} =$

**Table 1**

| $J_{\text{orb}}$, erg s | $8.81 \times 10^{49}$ | $6.17 \times 10^{49}$ | $3.17 \times 10^{49}$ |
|---|---|---|---|
| $j_0$ | 1.00 | 0.700 | 0.360 |
| $J_{\text{ac}}$, erg s | $1.72 \times 10^{49}$ | $1.45 \times 10^{49}$ | $1.25 \times 10^{49}$ |
| $J_{\text{ac}}/J_{\text{orb}}$ | 0.195 | 0.235 | 0.394 |
| $M_1, M\odot$ | 0.37 | 0.25 | 0.13 |
| $\delta_0$ | 0.206 | 0.139 | 0.0722 |
| $\Delta t_{\text{gr}}$, s | ∼0.04 | ∼0.2 | ∼10 |
| $\Delta t_{\text{ac}}$, s | ∼0.9 | ∼1.0 | ∼1.2 |

Note: $M_{\text{t}} = 1.8 M\odot$, $J_0 = 8.81 \times 10^{49}$ erg s, $t_{\text{grav}} = 4.7$ h.

4.7 h ≡ 16 920 s are given. In these calculations, we vary $j_0$, i.e., the fraction of the total angular momentum $J_0$ transformed into the orbital angular momentum $J_{\text{orb}}$. According to the data in Table 1, $j_0$ varies between 1 and 1/3. The next row gives the orbital angular momentum $J_{\text{ac}}$ at the time the low-mass NS fills its Roche lobe and the evolution of the binary with mass transfer begins. This evolutionary stage is called accretion for short; clearly, the solution of Eq. (2) with the second term $d\delta/dt \neq 0$ on its right-hand side corresponds to this stage. We see that $J_{\text{ac}}$ decreased significantly at the previous stage of NS approach under the effect of gravitational radiation alone: by a factor from ∼5 to ∼3, depending on $j_0$. Finally, the next rows give the sought-for values of the mass $M_1$ and $\delta_0$ (its ratio to the total mass $M_{\text{t}}$), which also decrease by a factor of almost 3. It is easy to see that the following strong inequality holds over the entire $j_0$ range: $M_1 \ll M_2$ (at least, $M_2$ is larger than $M_1$ by a factor of 4 for the first column).

Thus, our identification of the time interval between the two neutrino signals for SN 1987A with the time of approach of the components of the putative NS binary due to gravitational radiation alone (based on relation (3)) fits into the theory of the rotational mechanism of supernovae explosions for the following reason: The minimum NS mass $M_1 = 0.13 M\odot$ still exceeds the lower mass limit for stationary NS, $M_{1\min} = 0.095 M\odot$ (Blinnikov et al. 1990; Imshennik 1992; Aksenov et al. 1995), when the star explosively disintegrates with its transformation into an iron gas. The derived strong inequality $\delta_0 \ll 1$ qualitatively agrees with the first three-dimensional hydrodynamic calculations of the fragmentation of a rotating collapsar (Houser et al. 1994; Aksenov 1999), in which a ∼$0.1 M\odot$ mass ejection actually emerges.[1]

Nevertheless, an analysis of the physical effects disregarded in relation (3) is required to justify the results presented in Table 1 more reliably. To this end, the last two rows of Table 1 give $\Delta t_{\text{gr}}$ and $\Delta t_{\text{ac}}$ that have the following meaning: The values of these quantities should be compared with the $t_{\text{grav}} = 16\,920$ s specified above. Our estimates separately are smaller than the latter value at least by a factor of 2000 and are of no importance within the accuracy of the sought parameter $\delta_0$. The physical meaning of $\Delta t_{\text{gr}}$ is that, strictly speaking, from the time $t_{\text{grav}}$ in relation (3), we should subtract the time interval when the radius of the NS binary $a = a_{\text{R0}}$, where $a_{\text{R0}}$, the critical radius for filling the Roche lobe for a low-mass NS, is (Paczynski 1971)

$$a_{\text{R0}} = 2.16 R_{\text{NS1}} \delta_0^{-0.33}, \qquad (4)$$

where $R_{\text{NS1}}$ is the radius of a NS with mass $M_1$.[2] Thus, it follows from (4) that $a_{\text{R0}} \gg R_{\text{NS1}}$ at $\delta_0 \ll 1$. It is easy to show that $\Delta t_{\text{gr}} = t_{\text{grav}}(J_{\text{ac}}/J_{\text{orb}})^8$. The values of $\Delta t_{\text{gr}}$ are given in Table 1; they are very small. Imshennik and Popov (1998) analyzed in detail the mixed type of evolution described by the complete equation (2) with mass transfer (!) until $\delta$ decreased to its critical value of $\delta_{\text{cr}} = M_{1\min}/M_{\text{t}} = 0.053$, i.e., until the explosion time of the low-mass NS. By definition, mass transfer is possible if the condition for the immersion of a low-mass NS in its Roche lobe is satisfied, i.e., $a \leq a_{\text{R}}$, which includes the current critical radius for filling the Roche lobe,

$$a_{\text{R}} = 2.16 R_{\text{NS}} \delta^{-0.33}, \qquad (4')$$

obtained by a natural generalization of relation (4), depending on the parameter $\delta \leq \delta_0$ and the radius $R_{\text{NS}}$ of an NS with mass $M_1 = \delta M_{\text{t}}$. This inequality is actually satisfied throughout the mixed evolutionary stage of the NS binary, which can also be proved from physical considerations (Imshennik and Popov 1996).

---

[1] Interestingly, if a rotating collapsar fragmented into pieces of equal mass ($\delta_0 = 1/2$) at the same values of $m_{\text{t}} = 1.8$ and $j_0 = 1.0$, the time of approach of the components ($t'_{\text{grav}} \simeq 400$ s), according to formula (3), would be many times shorter than the time specified by the observations of the neutrino signals ($t_{\text{grav}} = 16\,920$ s)!

[2] As the NS radius $R_{\text{NS}}$, we may take interpolation formulas derived for numerical calculations of the radii of cold NS with the inclusion of low-mass NS, for examples, from Yaranowski and Krolak (1992). However, it should be remembered that, in our case, still very young and, hence, relatively hot NS whose radii are generally larger than the radii of cold NS are members of the binary. This implies that their Roche lobes are filled earlier and that their radii $a_{\text{R0}}$ are larger, but the related changed corrections $\Delta t_{\text{gr}}$ and $\Delta t_{\text{ac}}$ are still negligible compared to the sought-for time $t_{\text{grav}}$ from relation (3).

$\Delta t_{ac}$ follow from these calculations, because the calculations yield a time dependence $\Delta t(\delta)$ within the range $\delta_0 \geq \delta \geq \delta_{cr}$ with $\Delta t_{ac} = \Delta t(\delta_{cr})$. In contrast to $\Delta t_{gr}$, this quantity should be added to $t_{grav}$, because the two effects partly compensate each other. Therefore, their total effect, which is at a maximum in the last row of Table 1, is 8.8 s, which, we repeat, is negligible for the sought-for parameter $\delta_0$. In short, quantitatively including the second evolutionary stage of the components of the NS binary turned out to be completely unnecessary.

Nevertheless, it remains to determine how important the assumptions about the total mass of the Fe–O–C stellar core, $M_t$ (which was taken above to be $1.8 M_\odot$), about the total angular momentum $J_0$ (which was taken above to be $8.81 \times 10^{49}$ erg s), and, finally, about a zero orbital eccentricity $e_0 = 0$ are. At a fixed $t_{grav}$, obvious dependences follow from (3) for $\delta_0 \ll 1$: $(1 - \delta_0)\delta_0 \simeq \delta_0 \propto j_0^{0.89} m_t^{-1.7}$. In Table 1, the dependence $\delta_0 \propto j_0^{0.89}$, of course, holds. Since the decrease in initial $J_0$ as well as its increase are severely restricted by the condition of the dynamical rotational instability itself (Aksenov *et al.* 1995), the change in $M_1 \propto J_{orb}^{0.89} \propto J_0^{0.89}$ is small. The same is also true for the dependence $M_1 \propto M_t^{-1.7}$ that follows from the previous dependence of the dimensionless quantities, but for a different reason: the masses of Fe–O–C nuclei are limited to the standard range $1.2 M_\odot < M_t < 2 M_\odot$, as predicted by the stellar evolution theory. Note that, in this case, $M_t = 1.8 M_\odot$ in Table 1 is close to the upper limit of the range, so the possibility of its decrease in stellar cores before collapse definitely does not bring the corresponding values of $\delta_0$ outside the inequality $\delta_0 > \delta_{cr}$. We can reach the important conclusion that all the possible changes in parameters $M_t$ and $J_0$ do not qualitatively change the results that were formulated above based on the data of Table 1.

Next, let us consider not a circular but an eccentric orbit that may well result from the hydrodynamic fragmentation of a collapsar, i.e., with an initial eccentricity $e_0 \neq 0$ ($0 \leq e_0 < 1$). The problem of the evolution of an eccentric orbit for an NS binary was completely solved by Imshennik and Popov (1994). Although its solution (also analytical at an arbitrary value of $e_0$) is more complex than that in the limiting case of a circular orbit with $e_0 = 0$, the quantity $t_{grav}$ of interest can be expressed elegantly (see the pioneering paper by Peters 1964) at the same values of $M_t$ and $J_0$ as those in (3):

$$t_{grav} = 2.94 \times 10^{-4} \frac{j_0^8 \tau(e_0)}{m_t^{15} \delta_0^9 (1-\delta_0)^9} \text{ s}, \quad (5)$$

where the new (compared to (3)) dimensionless function $\tau(e_0)$ is given in the form of an easily calculated integral with a certain factor (both are functions of the parameter $e_0$):

$$\tau(e_0) = \frac{48}{19} \frac{1}{e_0^{48/19}(1 + \frac{121}{304}e_0^2)^{3480/2299}} \quad (6)$$

$$\times \int_0^{e_0} \frac{e^{29/19}(1 + \frac{121}{304}e^2)^{1181/2299}}{(1-e^2)^{3/2}} de.$$

This function $\tau(e_0) \geq 1$ differs only slightly from unity ($\tau(0) = 1$) as long as $e_0 \leq 0.5$ and increases steeply at $e_0 \to 1$, $\tau(e_0) \propto 1.81(1-e_0^2)^{-1/2}$. Our analysis shows that the limiting (singular) case of $e_0 = 1$ itself cannot be considered, because the necessary condition for the orbits being quasi-stationary is not satisfied. For this reason, this case has been excluded in the above $e_0$ range. The qualitative result obtained by Imshennik and Popov (1994) may be considered to be the establishment of such a rapid decrease in eccentricity $e$ ($de/dt < 0$ for any $e_0$ and $\delta_0$) due to gravitational radiation that $e = e_f \leq 0.1$ by the time the low-mass NS fills its Roche lobe, i.e., the orbit is almost indistinguishable from a circular orbit. To be more precise, this inequality also depends on the parameters $m_t$ and $j_0$ and on the radius $R_{NS1}$ (Imshennik and Popov 1994):

$$e_f < 0.044 \left(\frac{m_t^3}{j_0^2}\right)^{19/12} \left(\frac{R_{NS1}}{13.5 \text{ km}}\right)^{19/12}. \quad (7)$$

As our analysis indicates, the influence of these parameters and the radius is quantitatively small, so the previous eccentricity estimate is definitely justified by formula (7). This circumstance seems quite fortunate, because the entire theory of mass transfer was constructed precisely for circular orbits. It would be instructive to estimate the influence of a finite initial eccentricity $e_0$ on $\delta_0$. We take $j_0 = 1$ and $m_t = 0.9$ in formula (5) and substitute $e_0 = 0.9$, so the function $\tau(e_0) = 3$; hence we obtain $\delta_0 = 0.241$ at $t_{grav} = 4.7$ h, which should be compared with $\delta_0 = 0.206$ (the first column in Table 1). The following general conclusion can be reached: the influence of a finite initial eccentricity, $e_0 \neq 0$, is negligible and does not lead (like the influence of $m_t$) to any violation of the condition $\delta_0 > \delta_{cr}$. The latter inequality ensures the existence of a low-mass ejection ($\delta_0 \ll 1$) in the form of an NS.

One of the important results of the analytical model for the evolution of a close NS binary is the conclusion that, in the solution of Eq. (2) with mass transfer, the approach of the components very soon gives way to their recession from one another, and the dependence on the initial arbitrary parameter $\delta_0$ of the binary virtually disappears. We are talking here about the sought-for dependence of $\delta$ on time $\Delta t$ during the mixed evolutionary stage under the action of both factors: gravitational radiation and mass transfer (accretion) (see Fig. 6 in the paper by Imshennik and Popov (1998) cited above). This circumstance is formally similar to what has been said above about the influence of the other initial parameter $e_0$ on the solution. Thus, during the evolution under consideration, the influence of both emerged arbitrary parameters, $\delta_0$ and $e_0$, essentially vanishes, and the evolution of a close NS binary finishes irrespective of the choice of these parameters. However, a third arbitrary parameter, $\Delta J = J_0 - J_{\text{orb}}$, appeared in the course of our theoretical analysis. This parameter may prove to be crucial: Will the evolution end with the merger or explosion of a low-mass NS that has overfilled its Roche lobe? As Imshennik and Popov (1998) showed, the concept of a (circular) orbit and the point approximation for the gravitational interaction of both NSs become meaningless at the very end of the evolution.[3] A numerical solution of the three-dimensional hydrodynamic problem is required. This solution is extremely complex if the effects of gravitational radiation, neutrino radiation, and the nonideal equation of state for these NS are included in the problem. Of course, this solution is of relevant interest, but we have to restrict our selves to reasonable (?) physical estimates before it can be implemented after the overcoming of fantastic difficulties. Until now, it has been assumed that the more massive component with mass $M_2 = (1 - \delta_0)M_{\text{t}}$ does not collaps for the second time during the fragmentation of the rotating collapsar due to the residual total angular momentum $\Delta J_0$. The first neutrino signal was produced only by the primary collapse of the Fe–O–C stellar core that led to the formation of this collapsar before its fragmentation. However, even during the evolution, once the orbital radius $a = a_{\text{R0}}$ was reached, the high-mass NS gradually (in a time $\Delta t_{\text{ac}} \sim 1$ s) increased its mass up to $M_2 = (1 - \delta_{\text{cr}})M_{\text{t}}$. However, the centrifugal forces attributable to the angular momentum $\Delta J$ could still reliably prevent its secondary collapse. Here, it is important to note that this angular momentum $\Delta J$ can even exceed $J_{\text{ac}}$: for example, for the second column of Table 1, $J_{\text{ac}} = 1.45 \times 10^{49}$ erg s, while the corresponding $\Delta J = 2.64 \times 10^{49}$ erg s. Using relations (1), but for the final state of a binary with the orbital angular momentum $J_{\text{ac}}$ given above, we can easily find that the parameters of this orbit are $a_{\text{f}} = 2.22 \times 10^{7}$ cm and $v_{\text{f}} = 3.28 \times 10^{9}$ cm s$^{-1}$, so the angular frequency of the binary is $\omega_{\text{f}} = v_{\text{f}}/a_{\text{f}} = 1.48 \times 10^{2}$ s$^{-1}$. On the other hand, the angular frequency of a high-mass NS with the above value of $M_{2\text{f}} = 1.70 M\odot$ is $\Omega_{\text{f}} \sim 10^{5}$ s$^{-1}$ for rigid rotation (overestimate) or $\Omega_{\text{f}} \sim 10^{3}$ s$^{-1}$ for a typical differential rotation law (underestimate).

In short, the strong inequality $\Omega_{\text{f}} \gg \omega_{\text{f}}$ holds. However, the corotation condition is valid in close binaries. According to this condition, the orbital angular momentum must be much larger than the spin angular momenta, because the frequencies of these rotations, $\omega_{\text{f}}$ and $\Omega_{\text{f}}$, are equalized by tidal forces.

Strictly speaking, this condition takes effect even at the very beginning of the mixed evolutionary stage, i.e., since the time the low-mass NS fills its Roche lobe. Thus, a corollary of the corotation condition is the transformation of spin angular momentum $\Delta J$ into the orbital angular momentum of the binary. In this way, we also remove this main obstacle to the secondary collapse of a high-mass NS. The following legitimate question arises: Why did not the corotation condition prevent the appearance of a significant residual angular momentum $\Delta J$ during the fragmentation of the rotating collapsar? A completely justified answer to this question will probably be given after the construction of a three-dimensional (!) hydrodynamic model with a consistent allowance for the same physical factors that we mentioned when discussing the end of the evolution of a NS binary, but we can make a qualitative estimate. To be more precise, we can draw attention to the characteristic times of the hydrodynamic fragmentation, which was called hydrodynamic not by chance. These characteristic times are only $10^{-3}$ s, while the characteristic times of the mixed evolution for a NS binary, as has been repeatedly said above, are much larger, $\sim 1$ s. There is yet another factor in favor of the appearance of the residual angular momentum under consideration in a high-mass NS, as will be seen below.

Thus, a high-mass NS collapses in complete accordance with the standard model without rotation (Nadyozhin 1977a, 1977b, 1978). This collapse is accompanied by a standard neutrino signal, because during its evolution, the binary has virtually gotten rid of the bulk of its total angular momentum due to gravitational radiation. However, the remnant of the spin angular momentum was redistributed, according to the above assessment, to the orbital angular momentum, essentially transferring to the low-mass NS (see also below). It is important to recall that the integrated parameters of the standard neutrino signal are in satisfactory agreement with these parameters of the second neutrino signal from SN 1987A (Imshennik and Nadyozhin 1988; Blinnikov *et al.* 1988).

---

[3] Quantitatively, this last period of the evolution of a close NS binary is very short in terms of $\delta$; more specifically, it arises at $\delta = \delta^* = 0.082$, which exceeds the above critical value of $\delta_{\text{cr}} = 0.053$ only slightly. The parameter $\delta^*$ is virtually independent of the initial value of $\delta_0$, but depends on other parameters, $M_{\text{t}}$ and $J_{\text{orb}}$ (Imshennik and Popov 1998).

## 2. ESTIMATING THE (ELECTRON) NEUTRINO SPECTRA FOR SN 1987A

So far, we have been able to consistently analyze the collapse of a stellar Fe–O–C-core with initial rotation only in the one-dimensional, spherically symmetric statement of the problem by averaging the centrifugal force over the meridional angle (the only change of the problem then concerns the equation of motion from the complete system of four equations written in Lagrangian coordinates). The assumption that the local specific angular momentum $j$ is conserved played a key role in this great simplification of the problem. In this case, the centrifugal force $F_r$ was unambiguously expressed in terms of the initial rigid-rotation parameters $\omega_0 = $ const and $r_0(m)$ and the Eulerian radius $r = r(m,t)$, with $F_r \propto r^{-3}$. Obviously, the total angular momentum was then also automatically conserved in this case,

$$J_0 = \frac{2}{3} \int_0^{M_t} j\, dm = \frac{2}{3} \omega_0 \int_0^{M_t} r_0^2(m) dm,$$

with the factor $2/3$ being attributable precisely to the angular averaging of the centrifugal force:

$$F_r = \frac{2}{3}\frac{\omega^2 r^2}{r} = \frac{2}{3}\frac{\omega_0^2 r_0^4(m)}{r^3(m,t)}. \quad (8)$$

The first numerical calculations of this kind were performed by Imshennik and Nadyozhin (1977). The equations of state, the processes of material neutronization, and the description of neutrino processes did not differ in any way from their representation in the standard model. These hydrodynamic models, which were called quasi-one-dimensional because of the described allowance for the rotation effects, were computed until the formation of a hydrostatic equilibrium configuration that was called above a rotating collapsar.

Another study of the quasi-one-dimensional model with detailed analysis and testing of the results was carried out later (Imshennik and Nadyozhin 1992). In this paper, we focus on the discussion of the neutrino radiation parameters that were limited in the papers by Imshennik and Nadyozhin (1977, 1992) only to data on a light curve with a total energy $\mathcal{E}_{\nu\tilde{\nu}} = 3.3 \times 10^{52}$ erg (by the end of the computation $t = 2.9$ s) and to parameters of the neutrinosphere. Apart from the marked difference between the neutrino light curves for the calculations of the standard model ($\omega_0 = 0$) and the quasi-one-dimensional model with $\omega_0 = 0.86\Omega_0$ ($\Omega_0 = (GM_t/R_0^3)^{1/2}$ is the velocity at which the centrifugal and gravitational forces are equal on the surface of the stellar core near the equator),[4] an enormous difference appears between the neutrino optical depths indicated by several numbers in these light curves (see Fig. 6 from Imshennik and Nadyozhin 1992). Whereas $\tau \geq 100$ for the standard model starting from the maximum of the light curve, $\tau$ is typically several units (except the middle part of the light curve where it rises to 32) in the model with rotation. This implies that the approximation of radiative heat conduction used in the calculations is near its validity boundary for the quasi-one-dimensional model. In addition, the region inside the neutrinosphere encloses less than half of $M_t$ by mass (see Fig. 1 from the cited paper). By the end of the computation, this mass decreases by several more times ($\sim$5) (see Fig. 5 from the cited paper). We must also take into account the ambiguity in determining the neutrino optical depth itself (formally, four different determinations of it are possible!), in contrast to the photon optical depth. Nevertheless, an optically opaque region near the center of the collapsing stellar core since the time $t = 0.5$ s was introduced in the cited paper (see Fig. 1). The main objection to this introduction is as follows: The rotating collapsar in the initial two-dimensional axisymmetric configuration is a highly flattened structure (Aksenov *et al.* 1995) the polar radius of which is several times smaller than its equatorial radius. This flattening naturally entails a decrease in $\tau$ at least by the same number of times. In addition, it is clear that the rotating collapsar will be a hydrostatic equilibrium configuration only in an axisymmetric geometry. If the criterion for dynamical instability given above and obtained quantitatively, $\beta = 0.42 > 0.27 = \beta_{\rm cr}$ (Imshennik and Nadyozhin 1992), is satisfied, instability relative to the third (azimuthal) coordinate will rapidly (on the characteristic hydrodynamic time scale) transform it into a dumbbell-like configuration. The central region of the collapsar with the lowest specific angular momenta will be located in the bar of this dumbbell, which is very thin compared to the end balls. Here, it should be emphasized that, strictly speaking, the rotating collapsar cannot be formed as an equilibrium structure if the criterion $\beta > \beta_{\rm cr}$ is satisfied. It has the right to exist only for our idealization of axial symmetry and, of course, in terms of the quasi-one-dimensional model. Thus, the former center of the collapsing stellar core will certainly be seen not only toward the poles, but also on the sides of the bar, into which it falls during the nonlinear growth of dumbbell barlike ($m = 2$) instability (Tassoul 1978; Aksenov 1996).

---

[4] Since the adopted initial angular velocity $\omega_0$ is of great importance, we give its numerical value specified in the calculations of the quasi-one-dimensional model: $\Omega_0 = 1.78$ s$^{-1}$ (at $M_t = 2M\odot$ (!), $R_0 = 4.38 \times 10^8$ cm), and the corresponding value of $\omega_0 = 0.86\Omega_0 = 1.53$ s$^{-1}$ (the substitution of the coefficient 0.86 for 0.80 in the succeeding paper (Imshennik and Nadyozhin 1992) is among the corrections to the preceding paper (Imshennik and Nadyozhin 1977)

The above qualitative discussion gives us the right to advance the following hypothesis: the central region of the quasi-one-dimensional model with the formation of a rotating collapsar may be considered to be completely transparent for intrinsic neutrino radiation. Concurrently, it seems quite justifiable to retain the thermodynamic parameters in this region obtained in the calculations by Imshennik and Nadyozhin (1977, 1992) as moderately sensitive to this change of the status of neutrino radiation (from surface to bulk), because the optical depths are so small in these calculations. In any case, retaining such quantities as the central density and temperature ($\rho_c$ and $T_c$) we can be sure of their consistency, because they satisfy three of the four equations of the quasi-one-dimensional model (together with a violation of the entropy equation). Concurrently, we retain the total energy of the neutrino radiation, $\mathcal{E}^*_{\nu\tilde{\nu}} = 2.7 \mathcal{E}_{\nu\tilde{\nu}} = 8.9 \times 10^{52}$ erg. The numerical coefficient 2.7 is introduced by analogy with the standard model (Nadyozhin 1977a, 1977b, 1978), where the corresponding value of $\mathcal{E}_{\nu\tilde{\nu}} = 1.9 \times 10^{53}$ erg at time $t = 2.4$ s (see the table from Imshennik and Nadyozhin 1992) transforms into $\mathcal{E}^*_{\nu\tilde{\nu}} = 5.3 \times 10^{53}$ erg through justified estimates of the prolonged neutrino cooling stage of a hot NS (Imshennik and Nadyozhin 1988). The presented approach may be fraught with the overestimation of the bulk radiation parameters, particularly the total time of this process for an unstable rotating collapsar.[5] The interpretation of the observed time interval between the two neutrino signals from SN 1987A justified in the preceding section on the basis of the previously advanced rotational mechanism of the explosions of collapsing supernovae may serve as an indirect justification of the suggested approach.

Thus, we take the total energy carried away by neutrino radiation with allowance made for the cooling of the rotating collapsar (by the time $t_{\text{fin}} \simeq 6$ s),

$$\mathcal{E}^*_{\nu\tilde{\nu}} = 8.9 \times 10^{52} \text{ erg}. \quad (9)$$

At this time, the material is so cold that the neutrons degenerate (actually, this effect should have been taken into account at the end of the numerical calculation of the quasi-one-dimensional model, $t = 2.9$ s).

---

[5] We are grateful to D.K. Nadyozhin for a critical discussion in which he, in particular, drew special attention to the above circumstance.

Let us next consider the spectral properties of the neutrino radiation by assuming the bulk radiation mechanism due to the non-one-dimensional hydrodynamic collapse noted above. We restrict our analysis to the main reaction of the modified URCA-process $e^- + p \to n + \nu_e$, in which electron neutrinos $\nu_e$ are generated. If the material contains only free nucleons (see the estimates below), as was specified in the quasi-one-dimensional model, then, based on the Kirchhoff law and using the expression for the $\nu_e$ mean free path (Ivanova et al. 1969a; Imshennik and Nadyozhin 1972), we obtain the spectral specific radiation power:

$$\frac{q_\nu}{\rho} = \frac{4\pi B_\nu}{\rho} = \frac{1}{m_0} \frac{\ln 2}{(ft)_{np}} \frac{1}{1+\Theta} \quad (10)$$
$$\times \left(\frac{\varepsilon_\nu}{m_e c^2}\right)^5 \frac{1}{1 + \exp\left(\frac{\varepsilon_\nu}{kT} - \varphi\right)} \text{ g}^{-1}\text{s}^{-1},$$

where $\Theta = N_n/N_p$ is the degree of neutronization of the material, $\varphi$ is the dimensionless chemical potential of the electron gas (see below), $\varepsilon_\nu$ is the neutrino energy in erg, and the spectral power is given per unit interval of this energy. Below, we will be interested in the dependence of $(q_\nu/\rho)$ on energy $\varepsilon_\nu$, in which an allowance for the Pauli principle in the initial $\nu_e$ mean free path plays a crucial role: it is responsible for the exponential factor in (10). Note also that relation (10) gives the spectral power of the neutrino radiation $B_\nu$ per unit solid angle. By definition, the neutrino radiation in the reaction under consideration is isotropic in each (Lagrangian) particle of the material, so $B_\nu$ in (10) is also isotropic. The energy dependence of $(q_\nu/\rho)$ follows from (10):

$$\frac{q_\nu}{\rho} \propto \left(\frac{\varepsilon_\nu}{m_e c^2}\right)^5 \quad (11)$$
$$\times \frac{1}{1 + \exp\left(\frac{\varepsilon_\nu}{kT} - \varphi\right)} \propto \frac{x^5}{1 + \exp(x - \varphi)} = \phi(x, \varphi),$$

where $x = \varepsilon_\nu/kT$, $\varphi = \mu_e/kT$, and $\mu_e$ is the chemical potential of the electrons ($[\mu_e] = $ erg). The function $\phi = \phi(x, \varphi)$ with a specified parameter $\varphi$ has only one maximum with $x_{\max} > \varphi$ and the following asymptotics: $\sim x^5$ (for $x \to 0$) and $\sim x^5 e^{-x}$ (for $x \to \infty$). The quantity $x_{\max} = x_{\max}(\varphi)$ can be easily calculated and is given in Table 2. Curiously, at $\varphi = 10$, the exact solution for $x_{\max} = \varphi = 10$, while in the remaining cases of specified $\varphi$, $x_{\max}$ is given with four significant figures. We probably could have also calculated the $\nu_e$ energy averaged over spectrum (11), but it will suffice to understand that it is close to $x_{\max}$, and exceeds it slightly. These important quantities in the $\nu_e$ spectrum appear to have the following main property: $x_{\max} \geq 5$ for $\varphi \leq 5$; $x_{\max} \simeq \varphi$ for $\varphi \geq 5$). The inequality $x_{\max} \geq 5$ always holds for the lower limit of the maximum of the spectrum.

It would be appropriate to apply a correction for the self-absorption of neutrinos of sufficiently high energies in the layers surrounding the center of the rotating collapsar to the $\nu_e$ spectrum from (11). Let the mean optical depth of these layers be $\langle \tau_\nu \rangle = k$, where $k$ is an arbitrary number that may be close to unity ($k \leq 1$) even if the non-one-dimensional effects are taken into account. Of course, it is smaller than the optical depths obtained by Imshennik and Nadyozhin (1992) (see above). The value of $\langle \tau_\nu \rangle$ can be determined by the following estimate together with its spectral value:

$$\langle \tau_\nu \rangle = R \frac{\int_0^\infty B_\nu (1/l'_\nu) d\varepsilon_\nu}{\int_0^\infty B_\nu d\varepsilon_\nu}, \qquad \tau_\nu = \frac{R}{l'_\nu}, \qquad (12)$$

where $R$ is an effective radius (to simplify the estimation of (12), we disregard the weak dependence of the thermodynamic parameters in the surrounding layers on the current radius $r$). Thus, the mean $\langle \tau_\nu \rangle$ in (12) is weighted in radiation intensity $B_\nu$ from (10). Let us write the expression for the mean free path $l'_\nu$, which will be needed below (recall that it is uniquely related to (10) for $B_\nu$ by the Kirchhoff law):

$$l'_\nu = \frac{m_0}{\sigma_0} \frac{1+\Theta}{\Theta} \frac{1}{\rho} \left( \frac{m_e c^2}{\varepsilon_\nu} \right)^2 \frac{1 + \exp(\varepsilon_\nu/(kT) - \varphi)}{\exp(\varepsilon_\nu/(kT) - \varphi)}, \qquad (13)$$

whence we find an explicit expression for the mean $\langle \tau_\nu \rangle$:

$$\langle \tau_\nu \rangle = R \frac{\sigma_0}{m_0} \frac{\Theta}{1+\Theta} \rho \left( \frac{kT}{m_e c^2} \right)^2 \Psi(\varphi). \qquad (14)$$

The dimensionless function $\Psi(\varphi)$ is given by

$$\Psi(\varphi) = \left[ 7 \int_0^\infty \frac{x^6}{1 + \exp(x-\varphi)} dx \right] \qquad (15)$$

$$\times \left[ \int_0^\infty \frac{x^5}{1 + \exp(x-\varphi)} dx \right]^{-1}.$$

This expression can be easily tabulated by numerical integration. Table 2 gives the values of $\Psi(\varphi)$. In conclusion, we can easily determine the dimensionless energy $y = \varepsilon_\nu/(kT)$ that with $l'_\nu$ from (13) separates the spectrum into two parts: $\tau_\nu < k$ if $x < y$ and $\tau_\nu > k$ if $x > y$. To this end, we equate the spectral depth $\tau_\nu$ to the mean $\langle \tau_\nu \rangle$ from (14). Substituting Eq. (13) for $l'_\nu$ and Eq. (14) for $\langle \tau_\nu \rangle$ into the equality $\tau_\nu = \langle \tau_\nu \rangle$

**Table 2**

| $\varphi$ | 0 | 2.5 | 5.0 | 7.5 | 10.0 | 20 |
|---|---|---|---|---|---|---|
| $x_{\max}(\varphi)$ | 5.033 | 5.303 | 6.327 | 8.008 | 10.00 | 18.97 |
| $\Psi(\varphi)$ | 42.30 | 44.29 | 50.14 | 59.70 | 71.39 | 125.9 |
| $y(\varphi)$ | 6.509 | 6.750 | 7.396 | 8.757 | 10.57 | 19.32 |

yields a transcendental equation for the sought-for quantity $y$:

$$y^2 = \Psi(\varphi)[\exp(\varphi - y) + 1]. \qquad (16)$$

The numerical solution of Eq. (16) is also presented in Table 2. We see that $y$ exceeds $x_{\max}$ only slightly for all values of $\varphi$, implying that the $\nu_e$ spectrum at $k \simeq 1$ is cut off immediately after its maximum. For $k < 1$, the cutoff boundary is shifted slightly to higher energies. Note that the true value of $k$ can be determined in future hydrodynamic calculations, but its closeness to unity follows from the paper by Imshennik and Nadyozhin (1992).

Below, we apply these simple estimates to the thermodynamic parameters of the quasi-one-dimensional model for a rotating collapsar. Unfortunately, only some of the characteristic quantities, and only in the form of $\rho_c$ and $T_c$, the central density and temperature, but not the parameter $\varphi_c$, the central chemical potential of the electrons from (10), are at our disposal.[6] To calculate the corresponding value of $\varphi_c$, we will do the following. We will use the electrical neutrality condition for the material, which is naturally satisfied in the hydrodynamic calculations of collapse. It is greatly simplified in the ultrarelativistic case for a Fermi–Dirac electron gas:

$$\frac{8\pi}{3} \left( \frac{kT}{ch} \right)^3 (\varphi^3 + \pi^2 \varphi) = \frac{\rho}{m_0(1+\Theta)}. \qquad (17)$$

We can determine $\varphi$ of interest from this condition (in the form of a cubic equation) by using the given parameters $\rho$ and $T$. We know these parameters at the stellar center, $\rho_c$ and $T_c$, for the time $t = 2.9$ s at which the computation ends (Imshennik and Nadyozhin 1992; see the table there) and for the earlier time $t = 0.0$ s at which neutrino opacity sets in (Imshennik and Nadyozhin 1977). Two pairs of these parameters are presented in Table 3. However, as we see from condition (17), the parameter $\Theta$ should also be specified to solve the cubic equation. In Table 3, it is varied for the first pair of $\rho_c$ and $T_c$ over a wide range; the lowest value from the previous case will suffice for the second pair. In reality, the hydrodynamic calculations yield values of the material neutronization parameter $\Theta$ that do not exceed 100. Therefore, we conclude from the data of Table 3 that $\varphi_c \simeq 10$ may serve as a rough estimate. The corresponding energies, the mean energy and the energy at the maximum of the spectrum, in (11) are then found to be equal (see Table 2):

$$\langle \varepsilon_\nu \rangle \simeq \varepsilon_{\nu\max} \simeq kT_c \varphi_c \qquad (18)$$
$$\simeq (53.4\text{--}50.7) \text{ MeV} \simeq 50 \text{ MeV}.$$

---

[6]In both their publications of the calculations of the quasi-one-dimensional model, Imshennik and Nadyozhin (1977, 1992) focused on the rotation effects and the possibility of the transformation of collapse into an explosion and provided only minimum information concerning the integrated quantities of the light curve for neutrino radiation and its total energy.

**Table 3**

| $\rho_c = 2.6 \times 10^{14}$ g cm$^{-3}$,<br>$T_c = 6.2 \times 10^{10}$ K<br>($kT_c = 5.34$ MeV) | | | | $\rho_c = 1.70 \times 10^{13}$ g cm$^{-3}$,<br>$T_c = 5.89 \times 10^{10}$ K<br>($kT_c = 5.074$ MeV) | |
|---|---|---|---|---|---|
| $\Theta_c$ | 12.6 | 100 | 1000 | $\Theta_c$ | 12.6 |
| $\varphi_c$ | 25.7 | 13.0 | 5.63 | $\varphi_c$ | 10.6 |

Estimate (18) should be improved further, but the unusual hardness of the neutrino spectrum compared to the standard model of collapse without rotation is beyond question. We emphasize that the steep fall attributable to the $\nu_e$ self-absorption considered above (see Table 2) takes place immediately after the maximum of the spectrum.

Next, let us first estimate the total number of $\nu_e$ required for the almost complete neutronization of the material of a rotating collapsar ($\Theta \leq 100$) independent of the mean energy of the $\nu_e$ spectrum:

$$N_\nu = 1.8 M_\odot \frac{1}{m_0} \left( \frac{Z_{\text{Fe}}}{A_{\text{Fe}}} \right) = 1.0 \times 10^{57}.$$

On the other hand, according to the previous estimate of $N_\nu$, the total energy $\mathcal{E}_\nu$ of these neutrinos with $\langle \varepsilon_\nu \rangle = 50$ MeV is

$$\mathcal{E}_\nu = N_\nu \langle \varepsilon_\nu \rangle = 8.0 \times 10^{52} \text{ erg},$$

a value that is only slightly lower than $\mathcal{E}_{\nu\tilde{\nu}}$ from (9). To be more precise, about 10% of the energy remains for antineutrinos, $\tilde{\nu}_e$:

$$\mathcal{E}_{\tilde{\nu}} = \mathcal{E}_{\nu\tilde{\nu}} - \mathcal{E}_\nu = 0.9 \times 10^{52} \text{ erg}.$$

Note that the necessary condition $\mathcal{E}_{\nu\tilde{\nu}} > \mathcal{E}_\nu$ is still satisfied, with nothing remaining for muon and taon neutrinos ($\mathcal{E}_{\nu_\mu \tilde{\nu}_\mu} = \mathcal{E}_{\nu_\tau \tilde{\nu}_\tau} = 0$). Meanwhile, the large deficit of $\tilde{\nu}_e$ was clear from our estimate of the electron chemical potential. According to Imshennik and Nadyozhin (1972), the latter is related to the neutrino chemical potential by a simple relation for the thermodynamic equilibrium conditions of the $\nu_e(\tilde{\nu}_e)$ and $e^-(e^+)$ degenerate gases:

$$\psi = \varphi - \ln \Theta.$$

Thus, for $\varphi = 10$, we obtain $\psi = 4.6\text{--}7.7$, depending on $\Theta = 100\text{--}10$. This implies that the number of $\tilde{\nu}_e$ is strongly suppressed, being proportional to $\exp(-\psi)$, compared to the number of $\nu_e$.

It may seem that we use the parameters $T_c$ and $\rho_c$ noncritically, with the density of the material (on the order of nuclear density!) playing a major role in the achievement of high characteristic values of $\varphi_c \simeq 10$. However, the density $\rho_c$ was carefully analyzed by Imshennik and Nadyozhin (1992), including its dependence on the initial values of the rotation parameter $\omega_0$, using the derived analytical formula that relates this density to the relative centrifugal force near the center of the rotating collapsar. This analysis inspired confidence that initial rigid rotation with a sufficiently large total angular momentum, $J_0 \simeq 10^{50}$ erg s, does not prevent the emergence of a high density, $\rho_c \simeq 10^{14}$ g cm$^{-3}$, at the center of the rotating collapsar.

One would think that a significant admixture of iron nuclei etc. may be retained in the material under these conditions of relatively cold collapse (for a comparison of the central temperatures of the calculation under consideration and the standard model, see Imshennik and Nadyozhin (1992)). This admixture, of course, would immediately made condition (17), from which the chemical potential $\varphi$ was estimated, more complex. The iron mass fraction $X_{\text{Fe}}$ can be estimated using the paper by Imshennik and Nadyozhin (1965), where a universally accepted equation of state for the material was derived. Thus, using formula (29) from the cited paper for the specified parameters from Table 3, we find that $X_{\text{Fe}}/X_n \simeq 1.3 \times 10^{-2}$; i.e., for $X_n \simeq 1$, we actually obtain a negligible iron mass fraction, $X_{\text{Fe}} \simeq 10^{-2}$. Note that a correction on the order of unity to condition (17) then arises, but it also implies an increase in the sought-for parameter $\varphi$.

In conclusion, we may mention one subtle effect called the redistribution of specific angular momentum. In the quasi-one-dimensional model, this redistribution, of course, is constant, but it can change in principle during the fragmentation of a rotating collapsar. This would even be necessary at $\delta_0 \simeq 0.5$; i.e., when the collapsar is divided into equal pieces. What physical factors could provide this redistribution? First, we may point out the impact of three-dimensional tidal forces on any shear viscosity (for estimates of the neutrino viscosity, see Imshennik and Nadyozhin (1992)) and, finally, on the magnetic field. Of course, these factors should be included in

**Table 4**

| Model | $\mathcal{E}_1$, erg | $\mathcal{E}_2$, erg | $\mathcal{E}_3$, erg | $\bar{\varepsilon}_{\tilde{\nu}_e}$, MeV | $\bar{\varepsilon}_{\nu_e}$, MeV | $\bar{\varepsilon}_{\nu_{\mu,\tau}}$, MeV | $T$, s |
|---|---|---|---|---|---|---|---|
| I | $(3{-}14) \times 10^{53}$ | $(0.5{-}2.3) \times 10^{53}$ | $10^{52}$ | 12.6 | 10.5 | – | 20 |
| II | | | | 10 | 8 | 25 | 5 |

Note. $\mathcal{E}_1$ is the total burst energy transformed into neutrinos of all types; $\mathcal{E}_2$ is the total energy carried away by $\nu_i$, where $\nu_i = \nu_e, \tilde{\nu}_e, \nu_\mu, \tilde{\nu}_\mu, \nu_\tau, \tilde{\nu}_\tau$; $\mathcal{E}_3$ is the total energy carried away by $\nu_e$ during the neutronization of the star in a time $\sim 3 \times 10^{-2}$ s; $\bar{\varepsilon}_{\tilde{\nu}_e}$, $\bar{\varepsilon}_{\nu_e}$, $\bar{\varepsilon}_{\nu_{\mu,\tau}}$ are the spectrum-averaged energies of $\tilde{\nu}_e$, $\nu_e$, and $\nu_{\mu,\tau}$, respectively; $T$ is the duration of the neutrino burst.

future hydrodynamic models. At present, however, we may dispense with them if we are dealing with fragmentation into two pieces with a large difference in their masses ($\delta_0 \ll 1$). The initial rigid-rotation law then provides a high orbital angular momentum of a low-mass ejection almost without any redistribution of specific angular momentum. In our opinion, this circumstance is another argument for fragmentation with low values of the parameter $\delta_0$ (see Table 1) suggested in Section 2.

It would probably make sense to discuss the main conclusion of this section concerning the hard spectrum of electron neutrinos in the absence of other types of neutrinos from the viewpoint of possible future studies. We repeat that this conclusion requires a new series of quasi-hydrodynamic calculations of collapse in which, first, the bulk nature of the neutrino radiation in the generalized URCA process with corrections for self-absorption (Ivanova et al. 1969a) would be postulated, and, second, the more general kinetic equation for neutronization to which the complete system of kinetic equations reduces in nuclear statistical equilibrium (Imshennik and Nadyozhin 1982) would be used in the calculations of the degree of neutronization of the material instead of the rough approximation of kinetic equilibrium for the $\beta$-processes. We may take the risk of asserting that such calculations will generally lead to a decrease in the central temperature of the rotating collapsar, but with a simultaneous increase in its central density. According to relation (18), such changes in thermodynamic parameters will entail approximate conservation of or, most likely, a moderate decrease in the characteristic energies of the $\nu_e$ spectrum, which will probably be in better agreement with the LSD observations of the neutrino signal from SN 1987A than the very hard spectrum with $\langle \varepsilon_\nu \rangle \simeq \varepsilon_{\nu\max} \simeq 50$ MeV in (18) considered above.

## 3. A MODEL FOR A ROTATING COLLAPSAR AND A POSSIBLE INTERPRETATION OF THE EXPERIMENTAL RESULTS OBTAINED WITH NEUTRINO DETECTORS AT THE EXPLOSION TIME OF SN 1987A ON FEBRUARY 23, 1987

Let us consider how the various detectors operated during the explosion of SN 1987A could record the neutrino signals in terms of the model for a rotating collapsar that reduces to the following:

(1) Two neutrino bursts separated by a time $t^{\mathrm{grav}} \sim 5$ h must exist.

(2) The neutrino flux during the first burst consists of electron neutrinos ($\nu_e$) with a total energy $\mathcal{E}_{\nu_e} = \mathcal{E}_{\nu_e} = 8.9 \times 10^{52}$ erg; the neutrino energy spectrum $\phi(\mathcal{E}_{\nu_e}, \varphi)$ (11) is hard and asymmetric with mean energies in the range 25–50 MeV (Fig. 1); the duration of the neutrino radiation is $t_{\mathrm{fin}} \sim 2.9$–6 s.

(3) The second neutrino burst corresponds to the theory of standard collapse.

To compare the detector responses to these two neutrino bursts, we first give basic parameters of the neutrino fluxes obtained in the standard model and consider the neutrino detection methods.

The experimental searches for neutrino signals from collapsing stars began with the paper by Zeldovich and Guseinov (1965), who showed that gravitational collapse is accompanied by an intense short pulse of neutrino radiation. The search for collapse by detecting such a signal was first suggested by Domogatsky and Zatsepin (1965). The role of neutrinos in stellar collapses was considered by Arnett (1966), Ivanova et al. (1969b), Imshennik and Nadyozhin (1982), Nadyozhin and Otroshchenko (1982) [model I], Bowers and Wilson (1982), Wilson et al. (1986) [model II] and Bruenn (1987).

Parameters of the neutrino fluxes during the collapse of nonmagnetic, nonrotating, spherically symmetric stars were obtained by the above authors and are given in Table 4.

It follows from the theory of standard collapse that the total energy carried away by neutrinos of all types ($\nu_e, \tilde{\nu}_e, \nu_\mu, \tilde{\nu}_\mu, \nu_\tau, \tilde{\nu}_\tau$) corresponds to $\sim 0.1$ of the stellar core mass and is equally divided between these six components.

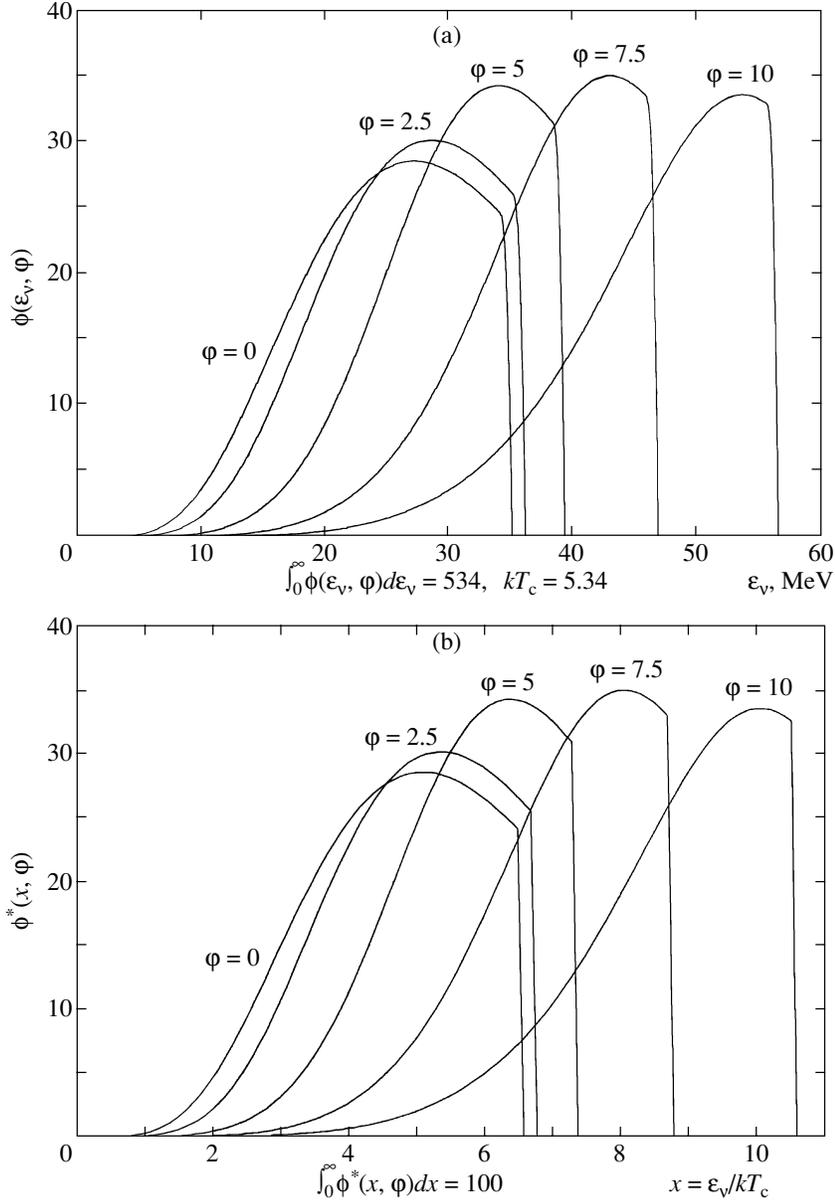

**Fig. 1.** Neutrino energy spectrum $\phi(\varepsilon_\nu, \varphi)$ (in arbitrary units) versus $\varepsilon_\nu$ in MeV (a) and $\phi^*(x, \varphi)$ (in arbitrary units) versus $x = \varepsilon_\nu/kT$ (b); $\varphi = \mu_e/kT$ is the dimensionless chemical potential of the electron gas.

Until now, Cherenkov ($H_2O$) and scintillation ($C_nH_{2n}$) detectors, which are capable of recording mainly $\tilde{\nu}_e$, have been used in searching for and detecting neutrino radiation. This choice is natural and is related to the large cross section for the interaction of $\tilde{\nu}_e$ with protons:

$$\tilde{\nu}_e + p \rightarrow e^+ + n, \quad (19)$$

$$\sigma_{\tilde{\nu}_e p} \cong 9.3 E_{e^+}^2 \times 10^{-44} \text{ cm}^2, \quad E_{e^+} \gg 0.5 \text{ MeV}.$$

Moreover, reaction (19) has a distinctive signature, and, as was first shown by Chudakov *et al.* (1973),

a proton may be used as a neutron catcher with the subsequent formation of deuterium ($d$) and the emission of a $\gamma$-ray photon with a time $\tau \approx 180-200$ $\mu$m:

$$n + p \rightarrow d + \gamma, \quad E_\gamma = 2.2 \text{ MeV}. \quad (20)$$

Electron antineutrinos $\tilde{\nu}_e$ can be detected by searching for a pair of pulses separated by the capture time. The first and second pulses are attributable to the detection of $e^+$ and $\gamma$-ray photons, respectively.

The $\nu_e$-scattering reactions have much smaller cross sections, but they allow the neutrino arrival direction to be determined in Cherenkov detectors:

$$\nu_e + e^- = \nu_e + e^-, \quad \sigma_{\nu_e e} = 9.4\varepsilon_{\nu_e} \times 10^{-45} \text{ cm}^2, \quad \varepsilon_{\nu_e} \geq 0.5 \text{ MeV}, \tag{21a}$$

$$\tilde{\nu}_e + e^- = \tilde{\nu}_e + e^-, \quad \sigma_{\tilde{\nu}_e e} = 3.9\varepsilon_{\nu_e} \times 10^{-45} \text{ cm}^2, \quad \varepsilon_{\tilde{\nu}_e} \geq 0.5 \text{ MeV}, \tag{21b}$$

$$\begin{cases} \nu_i + e^- = \nu_i + e^-, & \sigma_{\nu_i e} = 1.6\varepsilon_{\nu_i} \times 10^{-45} \text{ cm}^2, \quad \varepsilon_{\nu_i} \geq 0.5 \text{ MeV}, \\ \tilde{\nu}_i + e^- = \tilde{\nu}_e + e^-, & \sigma_{\tilde{\nu}_i e} = 1.3\varepsilon_{\tilde{\nu}_i} \times 10^{-45} \text{ cm}^2, \quad \varepsilon_{\tilde{\nu}_i} \geq 0.5 \text{ MeV}, \end{cases} \quad i = \mu, \tau. \tag{21c}$$

In scintillation detectors, neutrinos with energies above the threshold ($E_{\text{thr}}$) also interact with carbon:

$$\nu_i + {}^{12}\text{C} \rightarrow {}^{12}\text{C}^* + \nu_i, \quad E_{\text{thr}} = 15.1 \text{ MeV}, \quad i = e, \mu, \tau, \tag{22a}$$
$$\hookrightarrow {}^{12}\text{C} + \gamma, \quad E_\gamma = 15.1 \text{ MeV},$$

$$\nu_e + {}^{12}\text{C} \rightarrow {}^{12}\text{N} + e^-, \quad E_{\text{thr}} = 17.34 \text{ MeV}, \tag{22b}$$
$$\hookrightarrow {}^{12}\text{C} + e^+ + \nu_e, \quad \tau = 15.9 \text{ ms},$$

$$\tilde{\nu}_e + {}^{12}\text{C} \rightarrow {}^{12}\text{B} + e^+, \quad E_{\text{thr}} = 14.4 \text{ MeV}, \tag{22c}$$
$$\hookrightarrow {}^{12}\text{C} + e^- + \tilde{\nu}_e, \quad \tau = 29.3 \text{ ms},$$

where $E_{\text{thr}}$ is the threshold energy of the reaction, and $\tau$ is the mean lifetime of the $^{12}$B and $^{12}$N isotopes.

The cross sections for reactions (22a)–(22c) were calculated by Donnely (1973), Fukugita *et al.* (1988), Mintz *et al.* (1989), Kolb *et al.* (1994), and Engel *et al.* (1996) and measured in the Los Alamos E225 (Athanassopoulos *et al.* 1990), LSND (Allen 1997), and KARMEN experiments (see, e.g., Mashuv 1998).

Reaction (22a), which yields a monochromatic line at 15.1 MeV, allows one to measure the fluxes of $\nu_\mu$ and $\nu_\tau$ with energies twice the $\nu_e$ energy (Ryazhskaya and Ryasny 1992) for standard collapse and the total neutrino flux with energy $\mathcal{E}_\nu > 15.1$ MeV for any model of collapse. Reactions (22b) and (22c) in the standard scenario yield a smaller number of events without oscillations. With oscillations, the number of events increases and becomes comparable to the effect produced by reaction (22a) at equal neutrinosphere temperatures for neutrinos and antineutrinos, $T_{\tilde{\nu}_e}$ (Aglietta *et al.* 2002).

The cross section for the interaction of neutrinos with $^{16}$O nuclei in Cherenkov detectors $\sigma(\nu^{16}\text{O})$ for $\varepsilon_\nu \leq 25$ MeV is low compared to the cross section $\sigma(\nu^{12}\text{C})$ (Bugaev *et al.* 1979).

In Cherenkov and scintillation detectors, the emission from $e^+$, $e^-$, and photons is recorded by a photomultiplier. A neutrino burst is identified by the appearance of a series of pulses in the range of amplitudes from the detector energy threshold ($E_{thr}$) to 50 MeV in a time from several seconds to several tens of seconds, depending on the model of collapse.

Four detectors were operating during the explosion of SN 1987A: two scintillation (in USSR and Italy) and two Cherenkov (in USA and Japan). Parameters of the detectors are given in Table 5.

The effects from reactions (19) and (21a), (21c) expected in the model of standard stellar collapse in the Large Magellanic Cloud are shown in Table 6. We see that reaction (19) gives the largest contribution. It should be noted that all of the detectors except the LSD could record only $e^+$; i.e., there was no signature of the event by which reaction (19) could be unambiguously identified.

### Features of the Construction of Scintillation Detectors

Since we explore the possibility of detecting electron neutrinos by the nuclei of the detector materials, we will briefly consider the features of the construction of scintillation detectors, because some of them have not been reported previously.

The Baksan Underground Scintillation Telescope (BUST) (Alexeyev *et al.* 1979) is a cube with sides 14 m square. There are two horizontal planes inside the cube. 400 scintillation counters $0.7 \times 0.7 \times 0.3$ m$^3$ in size are located on each of the cube's planes (six outer and two inner). Each counter is watched by one photomultiplier. The spacing between the horizontal planes separated by an absorber is about 3.6 m; the thickness of the absorber is $\sim$170 g cm$^{-2}$, $\sim$20 g cm$^{-2}$ of which is made up of iron. The scintillator is made of white spirit (Voevodskii *et al.* 1970); the molecular composition is C$_n$H$_{2n}$, $\overline{n} = 10$. The detection threshold of energy release in the module is $\sim$10 MeV. The working volume of the detector consists of counters located on three horizontal planes: the two inner and one lower; the scintillator mass is 132 t. In analyzing the results of February 23, 1987, the authors also used data from some of the outer detectors, which increased working mass to 200 t.

Table 5

| Detector | Depth of water equivalent, m | Working mass, t | Material | Detection threshold, MeV | Detection efficiency | | Background pulse frequency $m$, s$^{-1}$** |
|---|---|---|---|---|---|---|---|
| | | | | | $e^+$ spectrum of reaction $\tilde{\nu}_e p \to e^+ n$ (19) | $e^-$ spectrum of reaction $\nu_i e^- \to \nu_i e^-$ (21a, 21c)* | |
| BUST, USSR | 850 | 130(200) | $C_n H_{2n}$ | 10 | 0.6 | 0.15 (0.54) | 0.013 (0.033) |
| | | 160 | Fe | | | | |
| LSD, USSR–Italy | 5200 | 90 | $C_n H_{2n}$ | 5–7 | 0.9 | 0.4 (0.7) | 0.01 |
| | | 200 | Fe | | | | |
| KII, Japan–USA | 2700 | 2140 | $H_2O$ | 7–14 | 0.7 | 0.17 (0.54) | 0.022 |
| IMB, USA | 1570 | 5000 | $H_2O$ | 20–50 | 0.1 | 0.02 (0.18) | $3.5 \times 10^{-6}$ |

\* The detection efficiencies of the electron spectrum produced in the reactions $\nu_{\mu,\tau}(\tilde{\nu}_{\mu,\tau}) + e^- \to \nu_{\mu,\tau}(\tilde{\nu}_{\mu,\tau}) + e^-$ (21c) are given in parentheses.
\*\* The background is given in the energy range $E_{thr}$–50 MeV; for the Cherenkov detectors, the background is given for the recording of internal events.

The LSD detector (Badino et al. 1984; Dadykin 1979) operated under Mont Blanc at a depth of 5200 m.w.e. consists of nine modules located on three floors with an area of $6.4 \times 7.4$ m$^2$. The LSD height was 4.5 m. The module is an iron container with an area of $6.4 \times 2.14$ m$^2$, a height of 1.5 m, and a wall thickness of 2 cm, with eight cells separated by 2-cm-thick iron sheets in which $1 \times 1.5 \times 1$ m$^3$ scintillation counters are located. In fact, LSD is an iron scintillation detector. To reduce the influence of radioactivity from the surrounding rocks, the detector is shielded by steel plates. The total iron mass is about 200 t. The mass of the scintillator is made up of white spirit (Voevodskiĭ et al. 1970) is 90 t. Each counter is watched by three photomultipliers (FEU-49B). The high sensitivity and low background of the detector allow both $e^+$ and $n$ particles in reaction (19) to be detected.

The parameters of the Kamiokande II and IMB detectors are well known (see Hirata et al. 1987; Bionta et al. 1987).

*Experimental Data*

To understand how the recorded events during the collapse of SN 1987A on February 23, 1987, fit into the scenario of a rotating collapsar, we present them in the chronological sequence of observation:

(1) February 23, 1987, 2:52:36.79 UT: LSD recorded a cluster of five pulses in real time; the estimated probability of this being mimicked by background radiation is low (less than once in three years). No such event had been observed in two years of LSD operation (1985–1987) (Aglietta et al. 1987; Dadykin et al. 1987, 1988). At present, we may add that the detector recorded no similar signal over its entire operation time until 1999. However, it is worth noting that the background conditions in 1988 were improved by an additional shield.

(2) February 23, 1987: information about the event measured by the LSD was transmitted to the group

Table 6

| Detector | $K_{e^+}$ (19) | $K_{e^-}$ (21a) + (21c) | $K_{e^-}$ (21c) |
|---|---|---|---|
| LSD | 1.5 | 0.043 | 0.024 |
| BUST | 2 | 0.052 | 0.036 |
| KII | 17 | 0.53 | 0.36 |
| IMB | 6 | 0.4 | 0.35 |

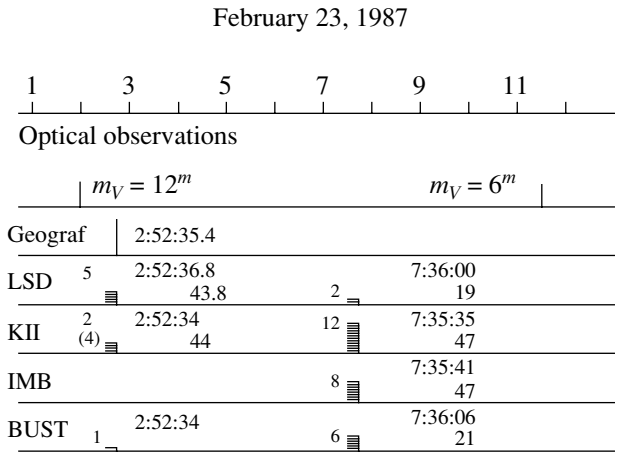

**Fig. 2.** Time sequence of the events recorded by various detectors on February 23, 1987.

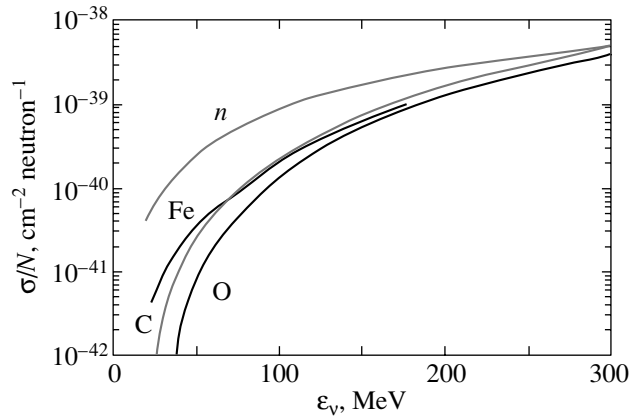

**Fig. 3.** Comparison of the reduced total cross sections with the neutrino cross section on a free neutron for the reaction $\nu_e + (A, Z) \to e^- + (A, Z+1)^*$ (Bugaev *et al.* 1979).

working with the GEOGRAV gravitational antenna (Rome).

(3) February 23, 1987, 10 : 40 UT: the first optical observation is made of SN 1987A in the Large Magellanic Cloud, ∼50 kpc from Earth, and is reported only the next day (IAU Circ. 1987).

It is worth emphasizing that the LSD event was observed and reported to Amaldi's group before information about the supernovae explosion was obtained. The data from the other detectors were analyzed later. Figure 2 shows the time sequence of the events recorded by various detectors on February 23, 1987. We see that there are two groups of events concentrated near the LSD time (2:52:36) and the IMB and KII time (7:35:35). All the events have been extensively discussed for the last several years.

Interestingly, strong correlations between the pulses from gravitational antennas and underground detectors were observed from 2:00 UT until 8:00 UT (Amaldi *et al.* 1987; Pizzella 1989; Aglietta *et al.* 1989, 1991), but the analysis of this fact is not the subject of this paper. Here, we wish to analyze the event recorded under Mont Blanc in terms of the model of a rotating collapsar. The events measured by underground detectors near 7:36 UT are interpreted by most physicists as the detection of antineutrino radiation, while attempts to explain the LSD data at 2:52 UT in a similar way run into difficulties.

Table 7 presents the parameters of the LSD pulses. The third column gives energies of the measured pulses corresponding to the two muon calibrations made before and after February 23, 1987 (each in a time interval of three months) and separated by two months. We see that the energy is determined with an accuracy of 20–25%.

*Difficulties in Interpreting the Effect Measured by LSD at 2:52 UT in the Case of Antineutrino Detection*

As was shown by Dadykin *et al.* (1989), a total neutrino radiation energy of $\mathcal{E}_\nu = 6\mathcal{E}_{\bar{\nu}_e} \approx 1.2 \times 10^{55}$ erg is required to explain the LSD effect in terms of the detection of antineutrino radiation if the conflicts with the results of other underground detectors have been eliminated. This energy is more than an order of magnitude higher than the binding energy of a neutron star with a baryon mass of about $2M_\odot$.

In addition, only one of the five measured pulses in the cluster was accompanied by a neutron-like pulse that was offset from the trigger signal in the detector by 278 μm and that had an energy of 1.4 MeV. Assuming the detection of five antineutrinos, one would expect the detection of five positrons accompanied, on average, by two neutrons. In this case, the probability of measuring one pulse attributable to the neutron capture by hydrogen and offset from the positron signal by $1.5\tau$, where $\tau$ is the neutron lifetime, is less than 5%.

**Table 7**

| Event no. | Time, UT ± 2 ms | Energy, MeV |
|---|---|---|
| 1 | 2:52:36.79 | 7–6.2 |
| 2 | 2:52:40.65 | 8–5.8 |
| 3 | 2:52:41.01 | 11–7.8 |
| 4 | 2:52:42.70 | 7–7.0 |
| 5 | 2:52:43.80 | 9–6.8 |
| 1 | 7:36:00.54 | 8 |
| 2 | 7:36:18.88 | 9 |

**Table 8**

| | | |
|---|---|---|
| F* | $\sigma = 1.27 \times 10^{-40}$ cm$^2$ | $E_{K,e^-}^{***} = 31.84$ MeV, |
| | | $E_\gamma = 1.82$ MeV, $n\gamma$: $\sum_n E_\gamma = 1.72$ MeV |
| GT** | $\sigma = 6.41 \times 10^{-41}$ cm$^2$ | $E_{K,e^-} = 30.84$ MeV, $E_\gamma = 1$ MeV, |
| | | $E_\gamma = 1.82$ MeV, $n\gamma$: $\sum_n E_\gamma = 1.72$ MeV |
| GT | $\sigma = 1.05 \times 10^{-40}$ cm$^2$ | $E_{K,e^-} = 27.84$ MeV, $E_\gamma = 4$ MeV, |
| | | $E_\gamma = 1.82$ МэВ, $n\gamma$: $\sum_n E_\gamma = 1.72$ MeV |
| GT | $\sigma = 1.27 \times 10^{-40}$ cm$^2$ | $E_{K,e^-} = 24.84$ MeV, $E_\gamma = 7$ MeV, |
| | | $E_\gamma = 1.82$ MeV, $n\gamma$: $\sum_n E_\gamma = 1.72$ MeV |

\* The Fermi level.
\*\* The Gamow–Teller resonance.
\*\*\* The electron kinetic energy.

The above arguments make the explanation of the LSD effect in terms of antineutrino detection implausible.

*A Possible Explanation of the LSD Effect in the Model of a Rotating Collapsar*

As was mentioned above, the collapsar emits electron neutrinos ($\nu_e$) with a total energy of $\mathcal{E}\nu_e^* \sim 8.9 \times 10^{52}$ erg, the spectrum shown in Fig. 1, and mean energies of 30–40 MeV for $\sim$2.9–6 s. These neutrinos can be recorded by the detector nuclei via the reactions

$$\begin{cases} \nu_e + (A,Z) \to e^- + (A, Z+1) \\ \nu_e + (A,Z) \to e^- + (A, Z+1)^*, \end{cases} \quad (23a)$$

$$\nu_e + (A,Z) \to \nu'_e + (A,Z)^*. \quad (23b)$$

The detectors operated on February 23, 1987, contained either oxygen, mainly $^{16}$O (KII and IMB), or carbon, mainly $^{12}$C, and iron $^{56}$Fe (LSD, BUST). It follows from the paper by Bugaev *et al.* (1979) that the reduced cross section $\sigma_{\nu_e n} = \sigma_{\nu_e A}/N$ for iron at $\varepsilon_\nu \leq 40$ MeV exceeds $\sigma_{\nu_e n}$ for oxygen by more than a factor of 20 ($\sigma_{\nu_e n}(^{56}\text{Fe}) > 20\sigma_{\nu_e n}(^{16}\text{O})$) (see Fig. 3). Thus, for these energies, the number of $\nu_e A$ interactions in the LSD (200 t of Fe) will be larger than that in the KII (1900 t $^{16}$O).

To answer the question of how the $\nu_e$Fe interactions are detected in the LSD, we turn to Table 8. As an illustration, this table presents the partial cross sections calculated for the following reaction for $\varepsilon_\nu = 40$ MeV (Gaponov *et al.* 2003):

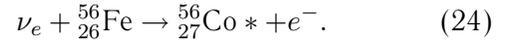
$$\nu_e + {}^{56}_{26}\text{Fe} \to {}^{56}_{27}\text{Co}* + e^-. \quad (24)$$

The ground level of $^{56}_{26}$Fe–$0^+$, the ground level of $^{56}_{27}$Co–$4^+$. The difference between the binding energies is $E[^{56}_{27}\text{Co}] - E[^{56}_{26}\text{Fe}] = 4.056$ MeV.

The threshold energy for reaction (24) is 8.16 MeV. An electron can be produced with an energy from $\sim$31.8 to 24.8 MeV, and its appearance is always accompanied by cascade $\gamma$-ray photons with a total energy of from 3.54 to 10.54 MeV. Recall that the critical energy ($\varepsilon$) in iron (the electron energy at which the ionization losses are equal to the radiative losses) is 21 MeV. Thus, an electron with an energy $E \geq \varepsilon$ in iron on a thickness of $d \geq 1$ t units (t is the radiation unit of length, 1 t units = 13.9 g cm$^{-2}$ (1.78 cm)) produces a small electromagnetic cascade. Calculations indicate that, during the interaction of electron neutrinos with $\varepsilon_{\nu_e} = 40$ MeV in a 2–3-cm-thick iron layer located between two scintillation layers, many more $\gamma$-ray photons than electrons fall into the scintillator (Dedenko and Fedunin 2003). The mean energies of these particles are $\sim$7–9 MeV. The detection efficiency of $\nu_e$Fe interactions ($\eta$) depends on the design of the detector and on the energy threshold: $\eta \sim 75\%$ for the inner part of the LSD ($\sim$90 t of Fe), $\eta \sim 35\%$ for its outer part ($\sim$110 t of Fe), and $\eta \leq 15\%$ for the BUST. We see that in the scenario under consideration, the mean energies recorded by a scintillation detector with an iron interlayer are close to the energies measured by the LSD on February 23, 1987. The estimated effect of the detection of neutrino radiation in the first phase of the collapse of a rotating star by different detectors is presented in Table 9. We used the cross sections of reactions (23a) and (23b) for $^{56}$Fe and $^{12}$C and the cross sections of reaction (23a) for $^{16}$O (Bugaev et al. 1978; Gaponov et al. 2003; Fukugita et al. 1988; Haxton 1987). The estimates were obtained for monochromatic neutrinos with $\varepsilon_{\nu_e} = 30$ MeV ($N_1$), $\varepsilon_{\nu_e} = 40$ MeV ($N_2$) and for the electron neutrino spectrum (Fig. 1) at $\varphi = 5$ ($N_3$) and $\varphi = 7.5$ ($N_4$).

**Table 9**

| Detector | Detection threshold | Estimated number of $\nu_e A$ interactions | | | | Estimated effect | Experiment |
|---|---|---|---|---|---|---|---|
| | | $N_1$ | $N_2$ | $N_3$ | $N_4$ | $N_2\eta$ | |
| LSD | 5–7 | 3.2 | 5.7 | 3.5 | 4.9 | 3.2 | 5 |
| KII | 7–14 | 0.9 | 3.1 | 1.2 | 2.5 | 2.7 | 2* |
| BUST | 10 | 2.8 | 5.2 | | | ~1 | 1** |

\* See De Rujula (1987a, 1987b).
\*\* Alexeyev et al. (1987).

We see that these estimates are consistent with the experimental data. If almost only electron neutrinos with a mean energy of ~35–40 MeV were emitted in the first collapse, then the experimental data correspond to the scenario for the rotational mechanism of a supernova explosion.

To be able to detect neutrino radiation from future stellar collapses (if the above scenario is realized), it is necessary to have detectors capable of recording not only $\tilde{\nu}_e$, but also $\nu_e$ with high efficiency. Two detectors of this type exist: the LVD (1.1 kt of scintillator, 1.1 kt of Fe) and the SNO (1 kt of D$_2$O). The Super K and Kamland detectors will be able to clearly see $\nu_e$ with $\varepsilon_{\nu_e} \geq 40$ MeV.

## CONCLUSIONS

We have put forward and partly justified our interpretation of the signal recorded by the underground LSD detector at 2:52 UT on February 23, 1987, as the first detection of a neutrino burst from the collapse of SN 1987A. We proceeded from the previously suggested rotational mechanism of the explosions of collapsing supernovae and the idea of taking into account the interaction of electron neutrinos with the nuclei of iron whose presence in the LSD's construction enormously increases the sensitivity to the first phase of collapse. A careful study of the corresponding nuclear reactions for the deneutronization and excitation of iron nuclides when electron neutrinos of sufficiently high energies, mainly from 20 to 50 MeV, interact with them led us to this conclusion. From the viewpoint of the rotational mechanism, it may be asserted that a rotating iron stellar core collapses in two phases separated by a relatively long time interval; the first phase of collapse was most likely detected by the LSD. According to this mechanism, the first phase is peculiar in that a rotating collapsar is formed in it with the emission of a very hard electron neutrino spectrum attributable to the reaction $e^- + p \rightarrow n + \nu_e$ with an energy at the maximum of the spectrum up to 50 MeV in the almost complete absence of electron antineutrinos and other types of neutrinos (muon and taon). A more detailed study of this spectrum based on numerical calculations (quasi-one-dimensional model) not only allows us to confirm the above properties but also to justify why other neutrino detectors have not recorded the first neutrino signal. We additionally took into account other possible nuclear reactions of electron neutrinos with oxygen nuclides in the KII and IMB Cherenkov detectors.

In addition, according to the same mechanism, we can reliably interpret the time interval between the first and second neutrino signals. This time interval is mainly attributable to the well-known gravitational radiation generated during the fragmentation of the rotating collapsar into a binary of NS with greatly differing masses. The evolution of such a binary was theoretically analyzed in detail in Section 1. At this time, gravitational radiation carried away a significant fraction of the initial angular momentum of the iron core, which was undoubtedly conserved during the first phase of collapse. It may be asserted that allowance for the rotation effects of the collapsing iron core alone allows us to theoretically interpret the two successive neutrino signals from SN 1987A. Until now, the presence of a neutrino signal on the LSD detector has been ignored in the theory of this famous event, which is besides enigmatically correlated with the signals from two gravitational antennas: in Italy and the United States. We have not yet been able to reasonably interpret how the mentioned antennas responded to gravitational radiation during the fragmentation of the rotating collapsar into a NS binary, which was theoretically estimated to be very modest.

It remains to be added that the second neutrino signal recorded by the KII, IMB, and BUST was associated with the detection of electron antineutrinos predicted in the universally accepted standard model for the secondary collapse of the high-mass NS in the putative binary once it had accreted the bulk of the mass from the low-mass NS and gotten rid of the remnants of its angular momentum. Theoretically, this signal corresponds to the standard hydrodynamic one-dimensional model of collapse without rotation with the formation of a neutrinosphere and with an equal energy distribution between all types of neutrinos. Electron neutrinos could not be recorded on the LSD detector at this time, because these neutrinos must have much lower mean energies, about 15 MeV. The cross section for their interaction with iron nuclides is too small.

Of crucial importance is the fact that a low-mass NS simultaneously ceases to exist once it has reached a critical mass of about $0.1 M_\odot$ and has been destroyed by an explosion relatively far (several hundred km) from the high-mass NS. This transformation into an ejection of iron with the release of recombination energy ($\sim 5$ MeV/nucleon) and in the presence of high kinetic energy ($\sim 0.3 \times 10^{51}$ erg) leads to an explosion with quite a sufficient total energy of about $10^{51}$ erg, and with a directed symmetry (the direction of the initial orbital motion of the ejection); i.e., it solves the fundamental problem of the transformation of iron-core collapse into a supernova explosion with an observed total energy of about $10^{51}$ erg.

## ACKNOWLEDGMENTS

Undoubtedly, we would be unable to reliably interpret the five famous events recorded by the LSD without the great progress in calculating the cross sections for the interaction of neutrinos with iron nuclides, or to reliably justify the high detection efficiency of neutrino interactions in the LSD without calculating the development of low-energy electromagnetic cascades. We wish to thank Yu.V. Gaponov, S.V. Semonov, L.G. Dedenko, and E.Yu. Fedunin, who made these calculations available to us. The strong stimulating influence of G.T. Zatsepin, whom we wish to thank, on the development of the theory of collapse and supernova explosion and the experimental methods for detecting neutrino radiation from these natural phenomena, including SN 1987A, is hard to overestimate. We wish to thank L.B. Okun for his interest in the problem, D.K. Nadyozhin for his helpful critical discussions, D.V. Popov for his help in calculating the evolution of a NS binary, and V.A. Matveev for his invariable interest in this problem. We are grateful to all members of the LSD collaboration, especially to C. Castagnoli, G. Cini-Castagnoli, V.L. Dadykin, V.G. Ryasny, O. Saavedra, and W. Fulgione for their fruitful discussions; to V.B. Braginskii, G. Pizzella, and E. Coccia for their discussion of the possibility of recording gravitational radiation during SN 1987A, and to N.Yu. Agafonova, N.A. Vulikh, and V.V. Kuznetsov for their help in preparing the paper for publication. This work was supported by the Russian Foundation for Basic Research (project nos. 03-02-16414a and 03-02-16436) and grants from scientific schools (projects NSh-1787.2003.2 and NSh-1782.2003.2).